\documentclass[preprint,pra,aps,superscriptaddress,showpacs,prd]{revtex4}
\usepackage{epsfig,amsmath,graphicx,amssymb}
\usepackage[colorlinks=fals,linkcolor=blue]{hyperref}
\def\be{\begin{equation}}
\def\ee{\end{equation}}
\def\bea{\begin{eqnarray}}
\def\eea{\end{eqnarray}}
\begin{document}
\title{Stability analysis of kinked DNA in $\mathcal{F}(K,\tau)$
model}
\author{Xiao-hua Zhou}
\email{xhzhou08@gmail.com} \affiliation{Department of Mathematics
and Physics, Fourth Military Medical University, Xi'an 710032,
People's Republic of China }

\date{\today}

\begin{abstract}
We phenomenologically analyze short DNA rings' stability by
discussing the second variation of its elastic free energy. Through
expanding the perturbation functions as Fourier series, we obtain
DNA rings' stability condition in a general case. By reviewing the
relationship between the Kirchhoff model and the worm-like road
chain (WLRC) model, we insert a spontaneous curvature term which can
partly reflect the twist angle's contribution to free energy in the
WLRC model and name this extended model the EWLRC model. By choosing
suitable spontaneous curvature, stability analysis in this model
provide us with some useful results which are consistent with the
experimental observations.

\end{abstract}

 \maketitle

%%%%%%%%%%%%%%%%%%%%%%%%%%%%%%%%%%%%%%%%%%%%%%%%%%%%%

\section{Introduction}
An important biological function for DNA loops is that it has
abundant deformations under different conditions. This deformations
are mostly due to the interaction between the DNA chain and the
environment. For instance, protein operation \cite{Luger}, ion
concentration \cite{Han-Nature,Han-PNAS} and temperature change
\cite{Richard} all can induce deformations. Since it has been found
that DNA shapes and deformations can be investigated by the
Kirchhoff elastic theory, much work have been done based on this
theory \cite{Benham1977,Bauer,Benham1983,Tanaka,Tsuru,
Wadati,Shi,Julicher,Starostin,Balaeff}. However, modern experimental
techniques, such as optical tweezers \cite{Smith}, micromanipulation
\cite{Strick,Williams,Yip} and other techniques
\cite{Unger,Cramer,Lionnet}, indicate that DNA has abundant mechanic
characteristics which cannot be explained by classical Kirchhoff
theory. For example, under different pulling force, DNA has
different stretching ability \cite{Smith}. Thus, several different
models have been suggested to describe DNA chain in different
conditions. For instance, the wormlike chain (WLC) model
\cite{Goldstein} was used to describe DNA under a small external
force. And the WLRC model \cite{Manning,Fain,Bouchiat} is
appropriate to describe DNA with its double-helix structure under a
moderate force. Further, following the increase of external force,
new energy terms are introduced \cite{Zhouhj-PRL}, and the results
are highly consistent with the experimental observations.

Based on the Kirchhoff theory, the free energy density of duplex DNA
chain can be written as
\bea\label{Kirchhoff}
\mathcal{F}=\frac{1}{2}a_1(K_1-\bar{K}_1)^2+\frac{1}{2}a_2(K_2-\bar{K}_2)^2+
             \frac{1}{2}a_3(K_3-\bar{K}_3)^2,
\eea
where
$K_1=-\textbf{x}_2\cdot(d\textbf{x}_3/ds),~K_2=\textbf{x}_1\cdot(d\textbf{x}_3/ds),
~K_3=\textbf{x}_2\cdot(d\textbf{x}_1/ds)$ and
\{$\textbf{x}_1,\textbf{x}_2,\textbf{x}_3$\} denotes the basis of
the local coordinates on DNA chain \cite{Manning, Tu}, and $s$ is
the arc length of DNA centreline, $a_1,~a_2$ and $a_3$ are
constants. $\bar{K}_1$ and $\bar{K}_2$ are spontaneous curvature and
$\bar{K}_3$ is spontaneous tension. Let $\phi=\phi(s)$ be the angle
between $\textbf{x}_1$ and the main normal $\textbf{n}$, we have
\bea\label{relate}
K_1=K\cos\phi,~K_2=K\sin\phi,~K_3=\tau+\dot{\phi},
\eea
where $K=K(s)$ and $\tau=\tau(s)$ are the curvature and torsion of
the DNA centreline. An overdot denotes a differential with respect
to $s$. Then the energy density in (\ref{Kirchhoff}) can be changed
into
\bea\label{energy}
\nonumber
 \mathcal{F} = \mathcal{F}(K,\tau,\phi,\dot{\phi})
            &=&\frac{1}{2}K^2(a_1\cos^2\phi+a_2\sin^2\phi)\\\nonumber
            & &-K(a_1\bar{K}_1\cos\phi+a_2\bar{K}_2\sin\phi)\\
            & &+\frac{1}{2}a_3(\tau+\dot{\phi}-\bar{K}_3)^2+constant,
\eea
 and the shape equations are shown in \cite{Tu}. Considering
 the energy density in (\ref{energy}) contains $\phi$, thus, any
 $\mathcal{F}(K,\tau)$ model, such as the Helfrich model
\cite{Zhao}, the WLRC model \cite{Manning,Fain,Bouchiat} and the WLC
model \cite{Goldstein}, cannot be equal to the classical Kirchhoff
 model, because they cannot give any information about the twist angle $\phi$.
 But specially choose $a_1=a_2=A$ and $a_3=C$, and let $\dot{\phi}=c=constant$, the energy density in (\ref{energy}) is
 reduced to
\bea
 \mathcal{F}&=&\frac{A}{2}K^2-t_0AK+\frac{C}{2}(\tau+c-\bar{K}_3)^2+constant,
\eea
where $t_0=\bar{K}_1\cos\phi+\bar{K}_2\sin\phi$. If $t_0\equiv0$, we
get the WLRC model. But in this paper we choose $t_0=constant$ and
$\dot{\phi}=c=constant$, and name this model the extended worm-like
road chain (EWLRC) model. We think the EWLRC model with
$t_0=constant$ is better than the WLRC model to a certain extent
because the arbitrary constant $t_0$ can partly reflect that the
twist angle $\phi$ and spontaneous curvatures $\bar{K}_1$ and
$\bar{K}_2$ influence DNA shapes. Then we choose energy density as
the form
\bea\label{EWLRC}
 \mathcal {F}=\frac{A}{2} (K-t_0)^2+\frac{C}{2}(\tau-\omega_0)^2,
\eea
where $\omega_0=\bar{K}_3-c$ and we also name it the spontaneous
tension. Clearly, the EWLRC model is different from the model
discussed in \cite{Zhouhj-JCP} by choosing $a_1=a_2,\bar{K}_2=0$ and
$\bar{K}_3=0$ in (\ref{Kirchhoff}).

By studying the second variation of DNA's free energy, Zhao \emph{et
al.} \cite{Zhao} found that the chiral term in energy function could
induce abundant shape transformations of DNA rings in the Helfrich
model, which is in good agreement with the experimental observations
in \cite{Han-Nature,Han-PNAS}. In \cite{Zhouhj-JCP}, Zhou \emph{et
al.} showed that the spontaneous curvature plays a significant role
in the stability of kinked DNA. By choosing Euler angles as variable
for the free energy function, Panyukov and Rabin \cite{Panyukov}
generally investigated the effects of thermal fluctuations on
elastic rings and pointed out that spontaneous curvature is
important in affecting the spatial configurations of the ring. In
this paper, we discuss the second variation of the $\mathcal
{F}(K,\tau,\dot{K},\dot{\tau})$ model and give stability condition
of ring solution in Sec. \ref{sectII}. In Sec. \ref{sectIII}, we
take the EWLRC model as an example and obtain some useful results
which are very consistent with experimental observations. Finally, a
short discussion is given in Sec. \ref{sectIV}.

\section{Second variations of free energy and stability condition of DNA rings}\label{sectII}

We choose the free energy for the closed duplex DNA with the form
\cite{Capovilla,Thamwattana-AM}
\bea
F=\oint \mathcal {F}(K,\tau,\dot{K},\dot{\tau})ds+\lambda\oint ds,
\eea
where $\mathcal {F}$ is an arbitrary function and $ds$ is the
element of the arclength $s$, and $\lambda$ is the Lagrange
multiplier. The central axis $\textbf{R}(s)$ of duplex DNA chain
under small perturbations can be written as
\bea\label{eq5}
\textbf{R}^{'}(s)=\textbf{R}(s)+\psi(s)\textbf{n}+\varphi(s)\boldsymbol{\beta},
\eea
where $\psi(s)$ and $\varphi(s)$ are two small smooth functions, and
$\boldsymbol{\beta}$ and $\textbf{n}$ are the binormal vector and
the main normal vector, respectively. Under the perturbations, the
general shape equations of DNA are attained in \cite{Thamwattana-AM}
by studying $\delta^{(1)}F=0$. For the shape equations, a ring
solution with radii $R=1/K$ and $\tau=0$ induces
\bea\label{eq8}
\lambda=R^{-1}\mathcal {F}_{1}^0-\mathcal {F}^0.
\eea
Here we define $\mathcal {F}_{1}=\frac{\partial \mathcal
{F}}{\partial K},\mathcal {F}_{2}=\frac{\partial \mathcal
{F}}{\partial \tau},\mathcal {F}_{3}=\frac{\partial \mathcal
{F}}{\partial \dot{K}},\mathcal {F}_{4}=\frac{\partial \mathcal
{F}}{\partial \dot{\tau}}$, and an up note $(~)^0$ means in the case
$K=1/R$ and $\tau=0$, such as $\mathcal {F}_{1}^0=\mathcal
{F}_{1}|_{(K=1/R,\tau=0)}$.

 For different DNA models, they have different energy density functions :
$\mathcal {F}(K,\tau,\dot{K},\dot{\tau})$. For an arbitrary function
$\mathcal {F}$, we have
\bea\label{eq8}
  \delta^{(1)}\mathcal {F} &=& \delta^{(1)}K\mathcal{F}_{1}+\delta^{(1)}\tau\mathcal {F}_{2}+\delta^{(1)}\dot{K}\mathcal{F}_{3}+\delta^{(1)}\dot{\tau}\mathcal {F}_{4}, \\\nonumber
  \delta^{(2)}\mathcal {F} &=& \delta^{(2)}K\mathcal{F}_{1}+\delta^{(2)}\tau\mathcal{F}_{2}+\delta^{(2)}\dot{K}\mathcal{F}_{3}+\delta^{(2)}\dot{\tau}\mathcal{F}_{4}\\\nonumber
                           & &  +\delta^{(1)}K\delta^{(1)}\tau\mathcal{F}_{12}+\delta^{(1)}K\delta^{(1)}\dot{K}\mathcal{F}_{13}+\delta^{(1)}K\delta^{(1)}\dot{\tau}\mathcal{F}_{14}\\\nonumber
                           & & +\delta^{(1)}\tau\delta^{(1)}\dot{K}\mathcal{F}_{23}+\delta^{(1)}\tau\delta^{(1)}\dot{\tau}\mathcal{F}_{24}+\delta^{(1)}\dot{K}\delta^{(1)}\dot{\tau}\mathcal{F}_{34}\\
                           & & +\frac{1}{2}(\delta^{(1)}K)^2\mathcal{F}_{11}+\frac{1}{2}(\delta^{(1)}\tau)^2\mathcal{F}_{22}+\frac{1}{2}(\delta^{(1)}\dot{K})^2\mathcal{F}_{33}+\frac{1}{2}(\delta^{(1)}\dot{\tau})^2\mathcal{F}_{44}.
\eea
The first and second variations of total energy are
\bea\label{eq8}
  \delta^{(1)}F &=& \oint \Big[(\mathcal{F}+\lambda)\delta^{(1)}g^{\frac{1}{2}}+g^{\frac{1}{2}}\delta^{(1)}\mathcal{F}\Big]g^{\frac{-1}{2}}ds,\\
  \delta^{(2)}F &=& \oint \Big[(\mathcal{F}+\lambda)\delta^{(2)}g^{\frac{1}{2}}+g^{\frac{1}{2}}\delta^{(2)}\mathcal{F}+\delta^{(1)}\mathcal{F}\delta^{(1)}g^{\frac{1}{2}}\Big]g^{\frac{-1}{2}}ds.
\eea
To obtain $\delta^{(1)}F$ and $\delta^{(2)}F$, those variations:
$\delta^{(1)}K$, $\delta^{(2)}K$, $\delta^{(1)}\dot{K}$,
$\delta^{(2)}\dot{K}$, $\delta^{(1)}\tau$, $\delta^{(2)}\tau$,
$\delta^{(1)}\dot{\tau}$ and $\delta^{(2)}\dot{\tau}$ are
indispensable. We give a method in Appendix A to obtain those useful
terms.

For a planar ring with $K=1/R$ and $\tau=0$, we expand the two
arbitrary functions $\psi$ and $\varphi$ as the following Fourier
series
\bea
 \nonumber
     & & \psi=c_0+\sum^{\infty}_{n=1}c_n\cos(nKs)+\sum^{\infty}_{n=1}d_n\sin(nKs),\\\nonumber
     & & \varphi=e_0+\sum^{\infty}_{n=1}e_n\cos(nKs)+\sum^{\infty}_{n=1}h_n\sin(nKs),
\eea
and after a lengthy calculation, we get the second variation, it
reads
\bea\label{second}
 \nonumber
 \delta^{(2)} F&=&p_0+\sum^{\infty}_{n=1}\big[p_n( c_n^2+d_n^2)+q_n(e_n^2 + h_n^2) \\
                                  & &+ 2m_n(c_n h_n - d_n e_n)+2s_n(c_n e_n + d_n
                                  h_n)\big],
\eea
with
\bea
 \nonumber
   p_0 &=& \frac{\pi}{2} R^{-3}c_0^2\mathcal{F}_{11}^0,\\ \nonumber
   p_n &=& \frac{\pi}{2} R^{-3}(n^2-1)^2(n^2R^{-2}\mathcal{F}_{33}^0+\mathcal{F}_{11}^0),\\
   q_n &=& \frac{\pi}{2} R^{-2}n^2(n^2-1)\big[R^{-1}(n^2-1)(R^{-2}n^2\mathcal{F}_{44}^0+\mathcal{F}_{22}^0)+\mathcal{F}_{1}^0\big],\\ \nonumber
   m_n &=& \frac{\pi}{2} R^{-2}n(n^2-1)\big[R^{-1}(n^2-1)(R^{-2}n^2\mathcal{F}_{34}^0+\mathcal{F}_{12}^0)-\mathcal{F}_{2}^0\big],\\ \nonumber
   s_n &=& \frac{\pi}{2} R^{-4}n^2(n^2-1)^2(\mathcal{F}_{23}^0-\mathcal{F}_{14}^0).
\eea
It is easy to find that $n=1$ is corresponding to a trivial
translation. In the Eq. (\ref{second}), if we don't know whether
$p_n$ and $q_n$ are positive or not, it seems difficult to discuss $
\delta^{(2)} F>0$. Specially, if $p_n$ and $q_n$ are nonnegative,
such as the Helfrich model \cite{Zhao}, stability condition
$\delta^{(2)}F>0$ yields
\bea\label{condition}
p_n q_n-m_n^2-s_n^2>0.
\eea
For the Helfrich model and any $F(K,\tau)$ model, we have
$\mathcal{F}_{23}^0=\mathcal{F}_{14}^0=0$, then condition
(\ref{condition}) is reduced to $p_n q_n-m_n^2>0$. However, if $p_n$
and $q_n$ are nonnegative, only $2m_n(c_n h_n - d_n e_n)$ and
$2s_n(c_n e_n + d_n h_n)$ in (\ref{second}) have the possibility to
be negative. So any stable deformations need
\bea
c_n h_n - d_n e_n<0~or~c_n e_n + d_n h_n<0.
\eea
It indicates that any deformations in this case are nonplanar,
because we cannot choose $e_n=h_n=0$. Then, tangential and normal
deformations are coupled each other and they must occur
simultaneously. Only when $p_n<0$ in (\ref{second}) can induce
planar deformations ($c_n\neq0,d_n\neq0$ and $e_n=h_n=0$).

\section{The EWLRC model}\label{sectIII}

For the EWLRC model in (\ref{EWLRC}) we assume $A>0$ and $C>0$, then
the stability condition (\ref{condition}) is reduced to
\bea
\nonumber
\Delta&=&A(n^2-1)[A+C(n^2-1)]\\
      & &-A^2(n^2-1)t_0 R-C^2\omega_0^2R^2>0.
\eea
By solving the above inequality, one gets the stable range of $R$.
Here, we only show the upper boundary as an example
\bea\label{EWLRC-condition}
\nonumber
R&<&\frac{1}{2W^2\omega_0}\Big[\sqrt{(n^2-1)^2(t_0^2+4W^3)+4W^2(n^2-1)}\\
 & & +(n^2-1)\frac{t_0}{\omega_0}\Big],
\eea
where $W=C/A$ and we let $\omega_0>0,t_0>0$. For DNA rings with
fixed $R$, such as 168-bp DNA with $R\approx9.1$ nm, inequality
(\ref{EWLRC-condition}) implies that the lower states will be
instable and rings will change into kinked shapes if $t_0$
decreases. Let $t_0$ be negatively related to the concentration of
ion, this conclusion is consistent with the experimental results
that complex deformations will emerge following the increase of ion
concentration \cite{Han-PNAS}. However, we have reverse conclusion
for $\omega_0$ that it needs to be positively related to the
concentration of ion. Specially, if $\omega_0=0$, we need that $t_0$
is positively related to the concentration of ion (we will see it in
the late text).

 Considering the helix solution
$K=\frac{r_0}{r_0^2+h^2},\tau=\frac{h}{r_0^2+h^2}$, the shape
equations in \cite{Tu,Capovilla,Thamwattana-AM} are reduced to
($\lambda=0$)
\bea\label{helix}
\nonumber
   A[(r_0^2+h^2)t_0-r_0][r_0^2(1+r_0t_0)+h^2(r_0t_0-2)] & & \\
   +Cr_0[(r_0^2+h^2)\omega_0-h][(r_0^2+h^2)\omega_0+3h] &=& 0.
\eea
Choosing $A=50$ nm, $C=0.15$ nm and the values of $r_0,h,\omega_0$
in \cite{Thamwattana-PRE}, we obtain the corresponding $t_0$ shown
in Table \ref{tab1}. Specifically, letting $h=0,r_0=R$, we get the
equation for a ring
\bea\label{ring}
A(1-R^2t_0^2)-C R^2\omega_0^2=0.
\eea
Solving the above equation, we get $t_0$ in Table \ref{tab2} for
168-bp DNA and 126-bp DNA ($R\approx6.8$ nm). However, these
constants in Tables \ref{tab1} and \ref{tab2} only give us some
coarse values, because the experimental observations require that
$\omega_0$ and $t_0$ should relate to ion concentration. Actually,
using these constants is difficult to explain the phenomena that
126-bp circles in 1 mM Zn$^{2+}$ mostly are not kinked but 168-bp
circles mostly are kinked in the same condition \cite{Han-PNAS}.

\begin{table}
\caption{\label{tab1}The parameters for helix of A-, B-, and Z-DNA.
Here $r_0,~r_0/h$ and $\omega_0$ come from \cite{Thamwattana-PRE},
$t_0$ is obtained by solving Eq. (\ref{helix}).}
\begin{center}
\begin{tabular}{p{1.5cm}p{1.5cm}p{1.5cm}p{2cm}p{2.3cm}}
\hline \\[-1.8ex]\hline \\[-1.8ex]
DNA  & $r_0$ (nm)  &$r_0/h$& $\omega_0$ (nm$^{-1}$)& $t_0$ (nm$^{-1}$)\\[0.8ex]
\hline \\[-1.8ex]
  A-DNA & 1.3&  3.32 & 2.55 &0.69 or -0.56\\
  B-DNA & 1.0&  1.89 & 1.89 &0.77 or -0.33\\
  Z-DNA & 0.9&  1.24 & 1.38 &0.66 or 0.22\\
\hline \\[-1.8ex]\hline \\[-1.8ex]
\end{tabular}
\end{center}
\end{table}
\begin{table}
\caption{\label{tab2}The spontaneous curvature $t_0$ (nm$^{-1}$)
obtained by solving Eq. (\ref{ring}) of A-, B-, and Z-DNA. }
\begin{center}
\begin{tabular}{p{1.5cm}p{1.5cm}p{1.5cm}p{1.5cm}}
\hline \\[-1.8ex]\hline \\[-1.8ex]
DNA  & A-DNA  &B-DNA& Z-DNA\\[0.8ex]
\hline \\[-1.8ex]
  168-bp  & --& 0.037 & 0.080\\
  126-bp  & 0.046& 0.104 & 0.126\\
\hline \\[-1.8ex]\hline \\[-1.8ex]
\end{tabular}
\end{center}
\end{table}

\begin{figure}
\includegraphics[scale =0.7]{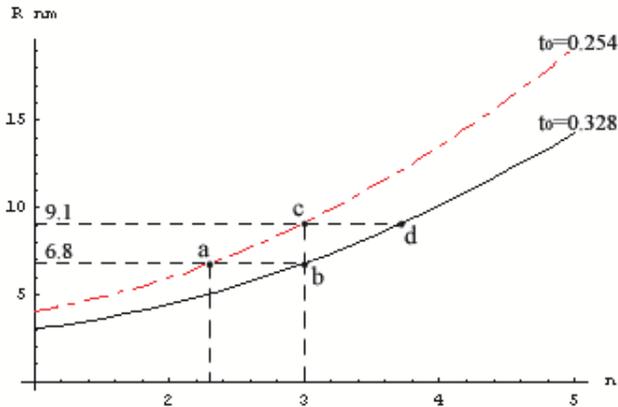}
\caption{\label{fig1}Stable ranges of DNA rings with
$W=0.154,~\omega_0=0$ and different $t_0$ (nm$^{-1}$). From a to b,
elliptic shapes will emerge for 126-bp DNA; from c to d, there are
trigonal shapes for 168-bp DNA, which is consistent with the value
that the number of kinks per circle is 3.83 \cite{Han-PNAS}.}
\end{figure}

Specially choosing $W=0.154$ and $\omega_0=0$ suggested by
Thamwattana \emph{et al} \cite{Thamwattana-PRE}, condition
(\ref{EWLRC-condition}) is reduced to
\bea\label{EWLRC-r}
R<\frac{(n^2-1)W+1}{t_0}.
\eea
Clearly, we need that $t_0$ is positively related to the
concentration of ion to explain that hight state deformation occurs
following the increase of ion concentration \cite{Han-PNAS}.
Moreover, with the same $W$ and $t_0$, Eq. (\ref{EWLRC-r}) indicates
that a big DNA ring will have higher state deformation than a small
one has. Supposing two rings with radius $R_1$ and $R_2$ ($R_1<R_2$)
are in the same condition (with the same $W$ and $t_0$), letting
$T(n)=(n^2-1)W+1$, we have
\bea\label{EWLRC-to}
\frac{T(n_2)}{R_2}<t_0<\frac{T(n_1)}{R_1},
\eea
where, $n_1$ and $n_2$ ($n_1\leq n_2$) lie on the deformation level
of the two rings with radius $R_1$ and $R_2$, respectively. Han
\emph{et al} \cite{Han-PNAS} find that 126-bp circles ($R_1=6.8nm$)
in 1 mM Zn$^{2+}$ mostly are not kinked and have somewhat elliptical
shapes ($n_1<3$) and 168-bp circles ($R_2=9.1nm$) in 1 mM Zn$^{2+}$
mostly are kinked, then we have
\bea
\frac{T(n_2)}{9.1}<t_0<\frac{T(3)}{6.8}.
\eea
We find that only $n_2=3$ can ensure $t_0$ is existent. So trigonal
deformation will emerge for 168-bp DNA, which is consistent with the
experimental result that the number of kinks is 3.83 per circler for
168-bp DNA in \cite{Han-PNAS}. Then, we get
$0.245~nm^{-1}<t_0<0.328~nm^{-1}$. Fig. \ref{fig1} shows the stable
range of two kinds of DNA rings with $W=0.154$ and $\omega_0=0$ in
(\ref{EWLRC-r}), from which we can see that trigonal deformation
will arise for 168-bp DNA rings and elliptic deformation will emerge
for 126-bp DNA rings.

In (\ref{EWLRC-to}) the existence of $t_0$ induces
\bea\label{EWLRC-w}
W[(n_2^2-1)R_1-(n_1^2-1)R_2]<R_2-R_1.
\eea
Choosing $R_1=6.8$ nm, $R_2=9.1$ nm and $n_1=3$, we find that the
above condition can always be satisfied with any $W$ (note we assume
$W>0$) when $n_2\leq 3$, because the left side of (\ref{EWLRC-w}) is
negative. This implies that 168-bp DNA rings with any $W$ will
change into trigonal shapes when 126-bp DNA rings change into
elliptic shapes in the same condition.

\section{Conclusions}\label{sectIV}

 We have generally investigated the second variation of DNA rings
 and shown the stability condition. However, only and if only $p_n\geq0$ and $q_n\geq0$
 condition (\ref{condition}) is valid. The EWLRC model and the
 Helfrich model are satisfied with this condition, and deformations in those models are must nonplanar.
If $\omega_0=0$, the stability condition (\ref{EWLRC-r}) needs that
$t_0$ is positively related to the ion concentration. Considering
$t_0=\bar{K}_1\cos\phi+\bar{K}_2\sin\phi$ and we choose
$\phi=constant$, which is consistent the conclusion in
\cite{Zhouhj-JCP} that the intercalation of Zn$^{2+}$ takes place
mainly at the exposed side of the DNA ring hence causing an increase
in its spontaneous curvature. In \cite{Han-PNAS} Han \emph{et al.}
find that there are somewhat elliptic shapes for 126-bp DNA and that
the number of kinks is 3.83 per circler for 168-bp DNA in 1 mM
Zn$^{2+}$, by choosing $W=0.154$, $\omega_0=0$ and
$0.245~nm^{-1}<t_0<0.328~nm^{-1}$ in the EWLRC model, our results
are very consistent with this phenomenon.

 \section*{Acknowledgements}

We thank Wang Jing for her helpful correction of our manuscript

\appendix{\textbf{Appendix A}}\label{appa}

  In this appendix, we present a method to obtain $\delta^{(1)}K$, $\delta^{(2)}K$, $\delta^{(1)}\dot{K}$,
$\delta^{(2)}\dot{K}$, $\delta^{(1)}\tau$, $\delta^{(2)}\tau$,
$\delta^{(1)}\dot{\tau}$ and $\delta^{(2)}\dot{\tau}$. For
simplicity, we only give the method and key steps. The variation of
$\textbf{R}(s)$ is
\bea
\delta\textbf{R}=\textbf{R}^{'}-\textbf{R}=\psi\textbf{n}+\varphi\boldsymbol{\beta}.
\eea
Let $g=\textbf{R}_{,x}\cdot\textbf{R}_{,x}$ ($
\textbf{R}_{,x}=\frac{d\textbf{R}}{dx}$, $x$ is the variable in
orthogonal coordinates), we have $ds=\sqrt{g}dx$ and
\bea
\nonumber
  \delta{g^{\frac{1}{2}}} &=&
                              -g^{\frac{1}{2}}K\psi+\frac{1}{2}g^{\frac{1}{2}}\big[\dot{\psi}^2+(\psi^2+\varphi^2)\tau^2+\dot{\varphi}^2\\\nonumber
                          & & +2(\psi\dot{\varphi}-\dot{\psi}\varphi)\tau\big]+\emph{O(3)},\\\nonumber
 \delta{g^{\frac{-1}{2}}} &=&
                               g^{\frac{-1}{2}}K\psi-\frac{1}{2}g^{\frac{-1}{2}}\big[\dot{\psi}^2+(\psi^2+\varphi^2)\tau^2+\dot{\varphi}^2\\
                          & &  -2\psi^2K^2+2(\psi\dot{\varphi}-\dot{\psi}\varphi)\tau\big] +\emph{O(3)}.
\eea
Where $\emph{O(3)}$ means the third and higher orders of $\psi$ and
$\varphi$. For an arbitrary function $\textbf{V}=\textbf{V}(s)$,
there is
\bea\label{dv}
 \delta{\dot{\textbf{V}}}= (1+g^{\frac{1}{2}}\delta g^{\frac{-1}{2}})(\delta\textbf{V})_{,s}+g^{\frac{1}{2}}\dot{\textbf{V}}\delta
 g^{\frac{-1}{2}}.
\eea
Using above equation, we have
\bea
\nonumber
 \delta{\dot{\textbf{R}}} &=& (1+g^{\frac{1}{2}}\delta g^{\frac{-1}{2}})(\delta\textbf{R})_{,s}+g^{\frac{1}{2}}\dot{\textbf{R}}\delta
                               g^{\frac{-1}{2}},\\
 \delta{\ddot{\textbf{R}}} &=& (1+g^{\frac{1}{2}}\delta g^{\frac{-1}{2}})(\delta\dot{\textbf{R}})_{,s}+g^{\frac{1}{2}}\ddot{\textbf{R}}\delta
                               g^{\frac{-1}{2}}.
\eea
Then, we get
\bea
 \delta{K^2}= 2\ddot{\textbf{R}}\cdot\delta\ddot{\textbf{R}}+\delta\ddot{\textbf{R}}\cdot\delta\ddot{\textbf{R}}.\nonumber
\eea
Consequently, we attain
\bea
\nonumber\label{dk}
 \delta K &=& \frac{1}{2}\left(K^2\right)^{-1/2}\delta
                K^2-\frac{1}{8}\left(K^2\right)^{-3/2}\left(\delta K^2\right)^2+\cdots\\\nonumber
         &=& (K^2-\tau^2)\psi-\dot{\tau}\varphi-2\tau\dot{\varphi}+\ddot{\psi}+K\Big[K^2\psi^2\\\nonumber
         & & -\frac{1}{2}\tau^2(\varphi^2+6\psi^2)-2\dot{\tau}\varphi\psi-6\tau\dot{\varphi}\psi-\dot{\varphi}^2\\\nonumber
         & & +\frac{1}{2}\dot{\psi}^2+2\ddot{\psi}\psi\Big]+\dot{K}\psi(\dot{\psi}-\tau\varphi)\\
         & & +\frac{1}{2}K^{-1}(\ddot{\varphi}+\dot{\tau}\psi+2\tau\dot{\psi}-\tau^2\varphi)^2+\emph{O(3)}.
\eea
Now, we have
\bea
  \delta K^{-1} = -K^{-2}\delta K+K^{-3}\left(\delta
  K\right)^2+\cdots.
\eea
The variation of $\textbf{n}$ is
\bea\label{dn}
  \delta{\textbf{n}} = K^{-1}\delta{\ddot{\textbf{R}}}+\ddot{\textbf{R}}\delta K^{-1}+\delta{\ddot{\textbf{R}}}\delta K^{-1}.
\eea
The variation of $\dot{\textbf{n}}$ can be obtained by
\bea\label{ddn}
 \delta{\dot{\textbf{n}}} = (1+g^{\frac{1}{2}}\delta g^{\frac{-1}{2}})(\delta\textbf{n})_{,s}+g^{\frac{1}{2}}\dot{\textbf{n}}\delta
                               g^{\frac{-1}{2}}.
\eea
 The first
variation of tension is {\cite{Zhouxh2564}}
\bea
\nonumber\label{dt1}
    \delta^{(1)}{\tau} &=& K(2\tau\psi+\dot{\varphi})+K^{-2}\dot{K}\big(\tau^2\varphi-\dot{\tau}\psi-2\tau\dot{\psi}-\ddot{\varphi}\big)\\
                       & & +K^{-1}\Big[\ddot{\tau}\psi+3\dot{\tau}\dot{\psi}-\tau^2\dot{\varphi}+\varphi^{(3)}+2\tau\big(\ddot{\psi}-\varphi\dot{\tau}\big)\Big].
\eea
Considering $\dot{\textbf{n}}\cdot\dot{\textbf{n}}=K^2+\tau^2$, we
have
\bea\label{dt2}
\delta^{(2)}{\tau} &=&
\frac{1}{2}\tau^{-1}\Big[2\dot{\textbf{n}}\cdot\delta^{(2)}\dot{\textbf{n}}+\left(\delta^{(1)}\dot{\textbf{n}}\right)^2-\delta^{(2)}K^2
-\left(\delta^{(1)}\tau\right)^2\Big].
\eea
Inserting (\ref{dk}), (\ref{ddn}) and (\ref{dt1}) into (\ref{dt2}),
after a lengthy calculation, we get
\bea
\nonumber
 \delta^{(2)}{\tau} &=& K^2\psi(3\tau\psi+2\dot{\varphi})-\tau^3(\psi^2+\varphi^2)-5\dot{\tau}\tau\psi\varphi\\\nonumber
                    & & +\tau^2(\dot{\psi}\varphi-5\psi\dot{\varphi})+2\tau(\dot{\psi}^2-\dot{\varphi}^2+3\ddot{\psi}\psi)\\\nonumber
                    & & +\dot{\tau}(8\dot{\psi}\psi-\dot{\varphi}\varphi)+2\ddot{\tau}\psi^2+2{\varphi}^{(3)}\psi+2\dot{\psi}\ddot{\varphi}\\\nonumber
                    & & +2\dot{\varphi}\ddot{\psi} +K^{-1}\Big[\ddot{K}\psi(\tau\psi+\dot{\varphi})+\dot{K}(\dot{\psi}\dot{\varphi}\\\nonumber
                    & & +\tau^2\psi\varphi)\Big]+K^{-2}\Big\{-\dot{K}^2\psi(\dot{\varphi}+\tau\psi)\\\nonumber
                    & & +\dot{\tau}^2\left[2\tau(\psi^2-\varphi^2)+3\psi\dot{\varphi}+3\dot{\psi}\varphi\right]\\\nonumber
                    & & +2\tau^3(\dot{\psi}^2-\dot{\varphi}^2+\ddot{\psi}\psi-\ddot{\varphi}\varphi-2\dot{\tau}\psi\varphi)\\\nonumber
                    & & +\tau^2(5\ddot{\psi}\dot{\varphi}+5\dot{\psi}\ddot{\varphi}+{\psi}^{(3)}\varphi+\psi{\varphi}^{(3)})\\\nonumber
                    & & +\dot{\tau}(3\ddot{\varphi}\dot{\varphi}-3\ddot{\psi}\dot{\psi}+{\varphi}^{(3)}\varphi-{\psi}^{(3)}\psi)\\\nonumber
                    & & +2\tau(\ddot{\varphi}^2-\ddot{\psi}^2+{\varphi}^{(3)}\dot{\varphi}-{\psi}^{(3)}\dot{\psi})\\\nonumber
                    & & +\tau^2\ddot{\tau}(\psi^2-\varphi^2)+8\tau^2\dot{\tau}(\dot{\psi}\psi-\dot{\varphi}\varphi)\\\nonumber
                    & & +2\ddot{\tau}\dot{\tau}\psi\varphi+4\dot{\tau}\tau(\ddot{\psi}\varphi+\psi\ddot{\varphi}+3\dot{\psi}\dot{\varphi})\\\nonumber
                    & & +\ddot{\tau}(\ddot{\varphi}\varphi-\ddot{\psi}\psi)+2\ddot{\tau}\tau(\psi\dot{\varphi}+\dot{\psi}\varphi)\\\nonumber
                    & & -{\psi}^{(3)}\ddot{\varphi}-\ddot{\psi}{\varphi}^{(3)}-\tau^4(\psi\dot{\varphi}+\dot{\psi}\varphi)\Big\}\\\nonumber
                    & & +2K^{-3}\dot{K}(\tau^2\psi+\dot{\tau}\varphi+2\tau\dot{\varphi}-\ddot{\psi})\\
                    & & \times(\tau^2\varphi-\dot{\tau}\psi-2\tau\dot{\psi}-\ddot{\varphi}).
\eea
Considering (\ref{dv}), we have
\bea
    \delta{\dot{K}} &=& (1+g^{\frac{1}{2}}\delta g^{\frac{-1}{2}})(\delta K)_{,s}+g^{\frac{1}{2}}\dot{K}\delta
                               g^{\frac{-1}{2}},\\
 \delta{\dot{\tau}} &=& (1+g^{\frac{1}{2}}\delta g^{\frac{-1}{2}})(\delta \tau)_{,s}+g^{\frac{1}{2}}\dot{\tau}\delta
                               g^{\frac{-1}{2}}.
\eea
Here, for simplicity, we only give $\delta \dot{K}$ and
$\delta{\dot{\tau}}$ for a planar ring with $K=1/R$ and $\tau=0$,
they are
\bea
     \delta^{(1)}{\dot{K}} &=& R^{-2}\dot{\psi}+\psi^{(3)}, \\
     \delta^{(2)}{\dot{K}} &=& 3R^{-3}\psi\dot{\psi}+R^{-1}(3\psi\psi^{(3)}
                               +3\dot{\psi}\ddot{\psi}-2\dot{\varphi}\ddot{\varphi})+R\ddot{\varphi}\varphi^{(3)}, \\
  \delta^{(1)}{\dot{\tau}} &=& R^{-1}\ddot{\varphi}+R\varphi^{(4)}, \\\nonumber
  \delta^{(2)}{\dot{\tau}} &=& R^{-2}(3\psi\ddot{\varphi}+2\dot{\psi}\dot{\varphi})-R^2(\ddot{\psi}\varphi^{(4)}+\ddot{\varphi}\psi^{(4)}+2\psi^{(3)}\varphi^{(3)})\\
                           & & +4\ddot{\psi}\ddot{\varphi}+4\dot{\psi}\varphi^{(3)}+3\psi\varphi^{(4)}+2\psi^{(3)}\dot{\varphi}.
\eea

\end{document}